%% file: main.tex
\documentclass[runningheads]{llncs}
\usepackage[T1]{fontenc}
\usepackage{subcaption}
\usepackage{graphicx}

\usepackage{hyperref} 
\newcommand{\reqref}[1]{\hyperlink{#1}{#1}}
\usepackage{comment}
\usepackage{amsmath}

\usepackage{rotating}
\usepackage{multirow}
\usepackage{booktabs}
\usepackage{tabularx}

\usepackage{xcolor}

\begin{document}
\title{
Containing the Reproducibility Gap: Automated Repository-Level Containerization for Scholarly Jupyter Notebooks
}
\titlerunning{Containing the Reproducibility Gap}
%
\author{Sheeba Samuel\inst{1}\orcidID{0000-0002-7981-8504} \and
Daniel Mietchen\inst{2}\orcidID{0000-0001-9488-1870} \and
Hemanta Lo\inst{1} \and
Martin Gaedke\inst{1}\orcidID{0000-0002-6729-2912}
}
\authorrunning{Samuel et al.}
%
\institute{Distributed and Self-organizing Systems, Chemnitz University of Technology, Chemnitz, Germany \\
\email{\{sheeba.samuel, martin.gaedke\}@informatik.tu-chemnitz.de}\\
\and
FIZ Karlsruhe — Leibniz Institute for Information Infrastructure, Berlin, Germany\\
\email{daniel.mietchen@fiz-karlsruhe.de}}

\maketitle              
\begin{abstract}
Computational reproducibility is fundamental to trustworthy science, yet remains difficult to achieve in practice across various research workflows, including Jupyter notebooks published alongside scholarly articles. Environment drift, undocumented dependencies and implicit execution assumptions frequently prevent independent re-execution of published research. Despite existing reproducibility guidelines, scalable and systematic infrastructure for automated assessment remains limited. We present an automated, web-oriented reproducibility engineering pipeline that reconstructs and evaluates repository-level execution environments for scholarly notebooks. The system performs dependency inference, automated container generation, and isolated execution to approximate the notebook’s original computational context. We evaluate the approach on 443 notebooks from 116 GitHub repositories referenced by publications in PubMed Central. Execution outcomes are classified into four categories: resolved environment failures, persistent logic or data errors, reproducibility drift, and container-induced regressions. Our results show that containerization resolves 66.7\% of prior dependency-related failures and substantially improves execution robustness. However, a significant reproducibility gap remains: 53.7\% of notebooks exhibit low output fidelity, largely due to persistent runtime failures and stochastic non-determinism. These findings indicate that standardized containerization is essential for computational stability but insufficient for full bit-wise reproducibility. The framework offers a scalable solution for researchers, editors, and archivists seeking systematic, automated assessment of computational artifacts.


\end{abstract}

\section{Introduction} 
Reproducibility is key to trustworthy, verifiable science \cite{national2019reproducibility,schnell2018reproducible}. For computational results, it means independent researchers can re-execute original code with the same data to obtain the same outputs under comparable conditions \cite{national2019reproducibility}. Despite strong policy support from journals and funding agencies \cite{hernandez2025reproducible,sayre2018the}, achieving reproducibility remains difficult in practice \cite{gelsleichter2025survey}.

A particularly visible and web-native manifestation of this challenge is the publication of computational artifacts as Jupyter notebooks \cite{kluyver2016jupyter}. They combine executable code, narrative explanations and visual outputs in a single, shareable artifact and are widely used across data-intensive disciplines \cite{badenhorst2019workflow,samuel2024computational}. Notebooks are frequently published alongside scholarly articles and hosted on web platforms such as GitHub, where they are expected to serve as executable companions to published results \cite{randles2017using}. However, notebook execution often depends on implicit and undocumented assumptions about software environments, library versions, data availability, execution order, and external services \cite{pimentel2019a,wang2020restoring}. These assumptions are rarely captured in machine-actionable form \cite{samuel2024computational}. Large-scale empirical studies consistently show that most published notebooks fail when re-executed in clean environments \cite{pimentel2019a,rule2018exploration,samuel2024computational,trisovic2022a}. 
Common failure modes include missing dependencies, incompatible library versions, unavailable datasets, hard-coded file paths, and reliance on external credentials \cite{hossain2025similarity,jiang2025exploring}. This reveals a persistent gap between reproducibility guidelines—emphasizing automation, environment specification, and transparency—and the reality of published computational artifacts \cite{hernandez2025reproducible}.
Existing approaches to notebook reproducibility fall into two main categories: large-scale measurement studies that diagnose execution failures \cite{pimentel2019a,rule2018exploration,samuel2024computational}, and execution-enabling tools such as containerization frameworks \cite{boettiger2015Docker}, configuration files, and interactive services like Binder. While diagnostic studies expose the scope of the problem, they offer limited remediation, and execution tools often require manual setup or project-specific customization, restricting their scalability across web-hosted research repositories \cite{hernandez2025reproducible}. In practice, computational environments are managed at the repository level rather than per notebook. However, environment reconstruction and evaluation remain largely ad hoc, and repository-level configurations are rarely integrated into automated workflows for systematic, large-scale reproducibility assessment \cite{gelsleichter2025survey}.

We treat computational reproducibility as an engineering problem requiring systematic automation, infrastructure, and empirical evaluation. We introduce an automated pipeline that operationalizes repository-level environment management for scholarly Jupyter notebooks. Rather than introducing new abstractions, the pipeline systematizes existing practices~-- including notebook-level dependency extraction from import statements and containerization~-- into an end-to-end process supporting automated execution, structured logging, output comparison, and large-scale analysis. We evaluate the pipeline on a corpus of real-world notebooks from GitHub referenced in PubMed Central \cite{samuel2024computational}.
This paper makes the following contributions:
\vspace{-4pt}
\begin{enumerate}
    \item \textbf{Automated End-to-End Reproducibility Pipeline.}
    We design and implement an automated, repository-centric execution infrastructure that crawls web-hosted scholarly repositories, performs static dependency inference, synthesizes container environments, and executes notebooks in isolation.
    \item \textbf{Infrastructure-Level Environment Reconstruction}
    We formalize repository-level environment reconstruction by combining declared dependencies with import-based inference and containerized isolation. 
    \item \textbf{Large-Scale Empirical Evaluation on Scholarly Web Artifacts.} We evaluate the pipeline on 443 notebooks from 116 GitHub repositories referenced in PubMed Central, providing quantitative evidence of improved execution robustness and systematically characterizing reproducibility outcomes across real-world research artifacts.
    \item \textbf{Structured Failure and Drift Analysis.}
    We provide fine-grained reproducibility metrics and categorize failure modes, offering actionable insights into the infrastructural roots of computational irreproducibility.
\end{enumerate}

\section{Related work}
Reproducibility of computational artifacts has been widely studied across software engineering \cite{jiang2025exploring}, scientific workflows \cite{badenhorst2019workflow,costa2025a,gruning2018practical}, and notebook-based research environments \cite{wang2020restoring}. Prior work has addressed environment capture \cite{chirigati2016reprozip,costa2025a}, dependency management \cite{boettiger2015Docker,gruning2018practical}, and large-scale empirical analysis \cite{rule2018exploration,pimentel2019a,samuel2024computational}; however, automated, repository-level, web-scale execution pipelines remain limited.

\textbf{Environment Capture and Container-Based Reproducibility}: Tools like ReproZip \cite{chirigati2016reprozip} bundle experiments by tracing system calls, yet focus on post hoc packaging rather than large-scale, automated repository reconstruction. Containerization has become a widely adopted strategy for reproducible research. Boettiger \cite{boettiger2015Docker} introduced early to Docker for reproducible research, showing how container images can preserve software environments and mitigate “dependency drift” in scientific workflows. Similarly, Singularity has been adopted in High-Performance Computing (HPC) environments where Docker may be restricted. In the life sciences, Grüning et al. \cite{gruning2018practical} emphasize practical reproducibility through containerized workflows and standardized environments. These works establish containerization as a mature infrastructure technology; however, they typically rely on manually curated images or workflow specifications.

The Binder ecosystem\footnote{\url{https://mybinder.org/}} and repo2docker\footnote{\url{https://github.com/jupyterhub/repo2docker}} facilitate interactive notebook execution by building containers from configuration files. While Binder lowers sharing barriers, it is designed for user interaction rather than structured reproducibility diagnostics, failure classification, or longitudinal analysis. Moreover, execution is typically notebook-level without systematic, fine-grained output comparison across re-executions.

Our work differs in two key aspects: (i) repository-level orchestration designed for scalable batch execution rather than interactive sessions, and (ii) structured logging and cell-level reproducibility metrics that enable quantitative assessment beyond mere executability.

\textbf{Dependency Management and Environment Reconstruction} Dependency management tools such as \href{https://conda.io/}{Conda} and pip-based workflows have been widely used to recreate Python environments. 
Rule et al. \cite{rule2018exploration} analyzed exploration and explanation patterns in computational notebooks, highlighting the hybrid nature of notebooks as both executable code and narrative artifacts. Pimentel et al. \cite{pimentel2019a} conducted a large-scale study using Conda environments based on declared dependencies or default Anaconda distributions, but lacked container isolation and import-based inference. More recently, Samuel et al. \cite{samuel2024computational} 
extended this analysis to PubMed Central-indexed repositories, identifying the absence of container-based execution and environment reconstruction as key limitations. Our work addresses these by introducing a web-scale, repository-level containerization pipeline designed to satisfy the following requirements:
\vspace{-4pt}
\begin{description}
    \item[\hypertarget{R1}{R1}] \textbf{Automated Repository Acquisition} 
    The system must automatically retrieve and process public GitHub repositories containing Jupyter notebooks.
    \item[\hypertarget{R2}{R2}] \textbf{Automated Environment Inference and Reconstruction}
    The system must reconstruct executable environments by extracting declared dependencies and statically analyzing notebook imports to infer missing packages.
    \item[\hypertarget{R3}{R3}] \textbf{Isolated Deterministic Execution}
    The system must execute computational artifacts within isolated container environments using structured logging to capture granular runtime metadata, including cell-level error traces and execution durations.    
    \item[\hypertarget{R4}{R4}] \textbf{Comparative Reproducibility Assessment and Reporting}
    The system must capture fine-grained execution logs, classify failure types, and compute quantitative cell-level reproducibility metrics, enabling transparent reporting and systematic evaluation beyond binary success measures
\end{description}
\vspace{-8pt}

\section{Methodology}
The core objective of our methodology is to transition from manual, error-prone notebook reproducibility checks to an automated, scalable and engineering-oriented evaluation pipeline. 
Our approach reconstructs the execution environment of each repository using containerization and re-executes all included notebooks in a clean-room setting. Reproducibility is then assessed by comparing original notebook outputs against those obtained from containerized re-execution, producing fine-grained, cell-level reproducibility metrics. Operating over public GitHub repositories, the pipeline is architected to systematically satisfy the system requirements \reqref{R1}-\reqref{R4} through a modular, four-stage process.

\begin{figure}
\centering
\includegraphics[width=0.8\textwidth, keepaspectratio]{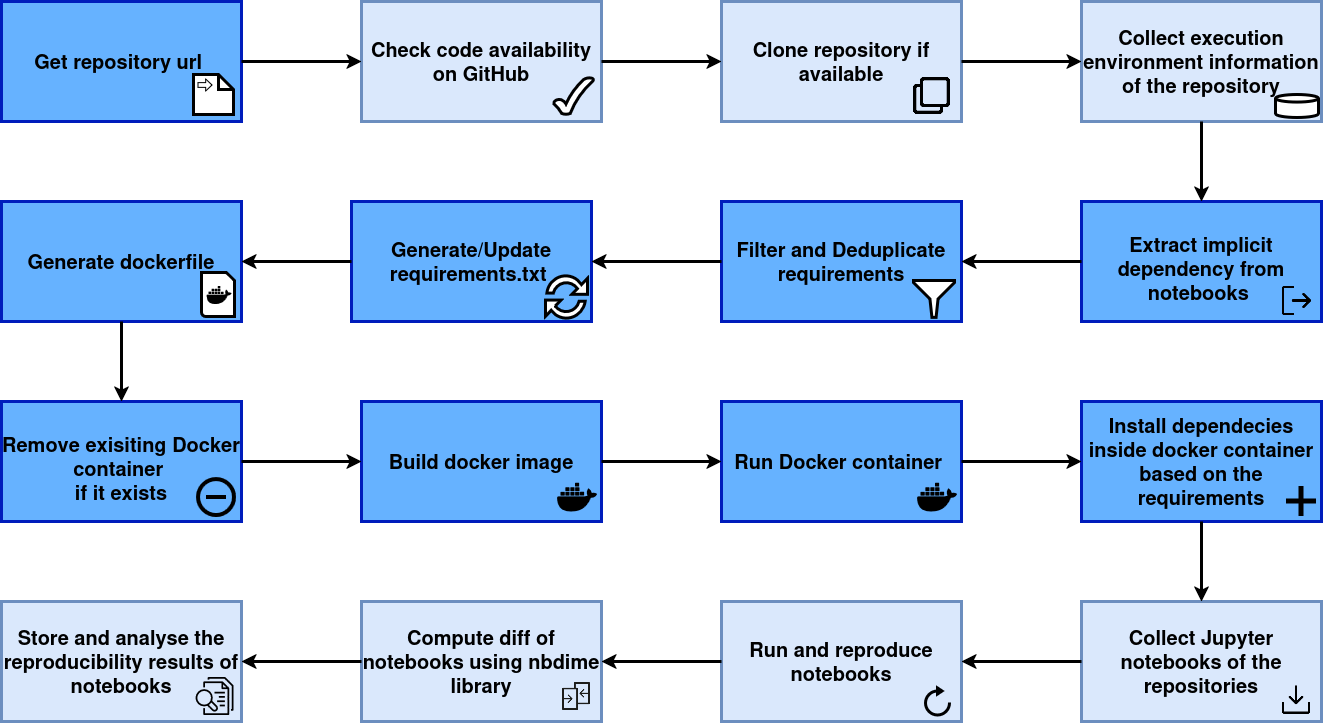}
\caption{Modified workflow from \cite{samuel2024computational}, adapted to incorporate automated environment reconstruction and containerized execution. Elements in dark blue signify newly integrated components, while light blue elements represent the original, unmodified process steps.
}
\label{fig:dockerworkflow}
\end{figure}

\subsection{System architecture and framework}
Figure \ref{fig:dockerworkflow} presents an overview of our end-to-end pipeline.
The framework is a modular, Bash-orchestrated pipeline integrating repository management, dependency extraction, Docker-based execution, and notebook reproducibility assessment. The architecture is guided by three core design principles: \textit{Isolation}, where each repository executes in an independent Docker container to eliminate host-system variability; \textit{Automation}, ensuring all pipeline steps occur without manual intervention for scalable experimentation; and \textit{Traceability}, where every execution, comparison, and failure is logged for auditing and longitudinal analysis. 

\subsection{Stages of the automated pipeline}
\label{sec:stages}
The pipeline follows a sequential process designed to satisfy the core requirements of automated discovery (\reqref{R1}), environment reconstruction (\reqref{R2}), isolated execution (\reqref{R3}), and reproducibility assessment reporting (\reqref{R4}).

\textbf{Stage 1: Repository Discovery, Validation, and Relational Tracking}  
This stage fulfills \reqref{R1} by implementing an automated discovery phase that validates repository accessibility to exclude removed or private projects. 
To support large-scale analysis, we maintain a normalized relational schema separating repository, notebook, and reproducibility data using unique \texttt{repository\_id}, \texttt{notebook\_id} and \texttt{run\_id} tags to ensure traceability. Each repository run represents a single containerized experiment.
Each notebook execution is stored in the \textit{notebook\_executions} table, including status, duration, code cell counts, and structured error metadata (type, category, message, cell index, error count). 
Reproducibility metrics~-- identical, different, and non-deterministic cell counts, cell index lists, and a reproducibility score (ratio of identical cells to total code cells)~-- are stored independently. 
We also track the markdown-to-code cell ratio to evaluate the impact of documentation within the dockerization pipeline. 

\textbf{Stage 2: Autonomous Environment Inference and Reconstruction }
To satisfy \reqref{R2}, the pipeline autonomously reconstructs repository-specific execution environments, which is a central contribution of our methodology. For each validated repository, the pipeline synthesizes a consolidated dependency specification through a multi-step process: explicit dependency discovery by detecting and parsing files like \texttt{requirements.txt} and \texttt{setup.py}; implicit dependency extraction via static analysis of all notebook import statements to capture undocumented requirements—a novel addition beyond baseline approaches; filtering and deduplication to remove standard library imports and minimize conflicts; and dynamic Dockerfile generation using a \texttt{python:3.10-slim} base image to pre-install all extracted dependencies and system utilities. The inferred dependencies are materialized into a synthesized \texttt{requirements.txt} file, which the generated Dockerfile utilizes to install the environment alongside essential tools like Jupyter and the \texttt{\href{https://github.com/jupyter/nbdime}{nbdime}}
library for notebook comparison. This automated synthesis avoids manual curation and mirrors the dependency resolution challenges practitioners face when reusing notebooks from public repositories.

\textbf{Stage 3: Containerized notebook execution}
To maintain a `clean-room'' execution environment, the pipeline includes a pre-execution cleanup step where any existing Docker containers or images associated with a specific repository are forcefully removed. This ensures that each experiment starts from a blank slate, preventing cross-contamination from previous failed builds or cached layers that might mask dependency conflicts. Fulfilling \reqref{R3}, the framework launches an isolated container for each repository and executes all notebooks via \texttt{jupyter nbconvert \mbox{-}\mbox{-}execute} with the \texttt{\mbox{-}\mbox{-}allow-errors} flag enabled. This ensures that notebooks producing partial results or runtime exceptions still generate output artifacts, allowing the framework to capture specific failure reasons (e.g.\ missing data files or incompatible library versions) rather than recording only binary failure outcomes. Executed notebooks are stored as separate output files, preserving successful outputs, execution metadata, and error traces. 

\textbf{Stage 4: Comparative Reproducibility Assessment and Reporting}
The final stage addresses \reqref{R4} by evaluating output fidelity through a structured comparison of original and container-executed notebooks using \texttt{nbdime}. The system computes the reproducibility score, the total code cells, reproducible cells, and non-reproducible cells. Results are persisted in a database linking identifiers to cell-level metrics. A structured logging directory captures Docker builds, dependency installation, and execution traces to enable root-cause failure analysis. 

\vspace{-10pt}
\section{Evaluation and results}
This section evaluates our container-based notebook reproducibility framework using repositories from our existing corpus \cite{samuel2023dataset,samuel2024computational}. We examine whether repository-level containerization enhances execution robustness and reproducibility of scholarly Jupyter notebooks compared to the previously deployed non-containerized baseline pipeline \cite{samuel2024computational}. We assess the proposed containerized pipeline as an automated infrastructure process and address three research questions:
\begin{itemize}
    \item \textbf{Execution robustness} (Validating \reqref{R2} \& \reqref{R3}): Does repository-level containerization improve successful execution rates versus the baseline pipeline?

    \item \textbf{Dependency handling } (Validating \reqref{R2}): How does explicit and implicit dependency specification affect execution outcomes?

    \item \textbf{Failure transparency} (Validating \reqref{R1} \& \reqref{R4}): Does isolated execution enable finer-grained reproducibility diagnostics?
\end{itemize}

\subsection{Experimental Setup}
\textbf{Corpus Selection}: Our evaluation builds directly on the original SQLite database used in \cite{samuel2023dataset,samuel2024computational}, which contains repository metadata, notebook identifiers and execution results obtained from prior non-containerized executions of 27,271 Jupyter notebooks from 2,660 GitHub repositories. This historical dataset serves as a baseline for comparative analysis. 

We selected 116 repositories with 443 Jupyter notebooks from our previous corpus. Following \cite{samuel2024computational}, we restrict analysis to Python-based Jupyter notebooks. To ensure a representative and diverse evaluation, we selected the first 100 repositories in execution order and augmented them with a random sample. The resulting dataset covers heterogeneous characteristics, including tutorial-style notebooks, Python versions 2.7 to 3.10, and varying reproducibility outcomes (identical and different results). While designed for web-scale execution, this evaluation uses a PubMed Central subset as a challenging testbed due to frequent dependency churn, implicit data assumptions, and limited documentation.

\textbf{Execution Environment} Experiments were conducted on a standardized Ubuntu 24.04 LTS environment with Docker 29.1.5, Python 3.12, 32GB RAM and 12-core CPU. This setup simulates an independent researcher’s attempt to reproduce results.

\textbf{Baseline Comparison} We evaluate our results against the non-containerized baseline established in our prior study \cite{samuel2024computational}, which relied on Conda environments derived solely from explicit dependency declarations. This enables direct assessment of the impact of repository-level isolation and automated environment inference on reproducibility outcomes.

\textbf{Metrics} We quantify the framework's performance using the following:
\begin{itemize}
\item \textbf{Dependency Installation Success}: The percentage of repositories where the environment reconstruction phase \reqref{R2} successfully resolves and installs all inferred dependencies.
\item \textbf{Success Rate}: A binary metric representing completion; a notebook is successful if it executes all cells to completion without encountering exceptions, crashes, or kernel interruptions.
\item \textbf{Reproducibility Score}: $\text{Score} = \frac{N_{identical\_cells}}{N_{total\_code\_cells}}$, measuring the output fidelity between the pipeline execution and the author's original notebook.
\item \textbf{Error Categorization}: A structured taxonomy used to log failure modes, including dependency, data, code, and logic errors.
\end{itemize}

\subsection{Results}
Table \ref{tab:stats} shows the statistics of the evaluation. To better understand the impact of repository-level containerization beyond aggregate metrics, we evaluate its impact through a comparative analysis of dependency resolution, execution success, output fidelity, and performance duration. Table \ref{tab:prev-vs-this} presents a comparative analysis of our containerized results against the baseline established in \cite{samuel2024computational}, mapping a subset of repositories across four distinct outcome classes: (i) environment-related failures resolved by containerization, (ii) persistent logic or data errors, (iii) reproducibility drift despite successful execution and (iv) regressions introduced by containerization. 

\begin{table}[h]
\centering
\caption{Summary of Dataset Composition and Execution Results. }
\label{tab:stats}
\begin{tabular}{lr}
\hline
\textbf{Metric} & \textbf{Count} \\ \hline
\textit{Dataset Composition} & \\
Total Repositories & 116 \\
Total Notebooks Identified & 443 \\ \hline
\textit{Environment \& Pre-execution} & \\
Successful Environment Provisioning (Docker) & 89 \\
Kernel not Found & 20 \\
Repositories with no python notebooks & 4 \\
Module not Found & 2 \\
Invalid Repository URLs* & 1 \\ 
Repositories with \texttt{requirements.txt} & 37 \\
Repositories without \texttt{requirements.txt} & 79 \\ \hline

\textit{Execution Outcomes} & \\
Total Notebooks Executed & 443 \\
\quad Notebooks Reproducible (Zero Errors) & 59 \\
\quad Notebooks Reproducible with Persistent Errors & 384 \\ \hline
\multicolumn{2}{l}{\small \textit{*Repository unavailable due to DMCA takedown}} \\\end{tabular}
\vspace{-10pt}
\end{table}

\textbf{Dependency Installation Improvement} 
The most significant engineering gain is observed in the transition of repositories from the baseline's failed states into Class (i).
Containerization successfully resolved dependency installation errors for 66.7\% (64 of 96) of the evaluated repositories. This shift from dependency failure to execution success highlights the pipeline's ability to stabilize environments that were previously not runnable. The remaining 33.3\% represent cases where issues beyond initial dependency installation persisted.

\begin{figure}
\centering
\includegraphics[width=\textwidth]{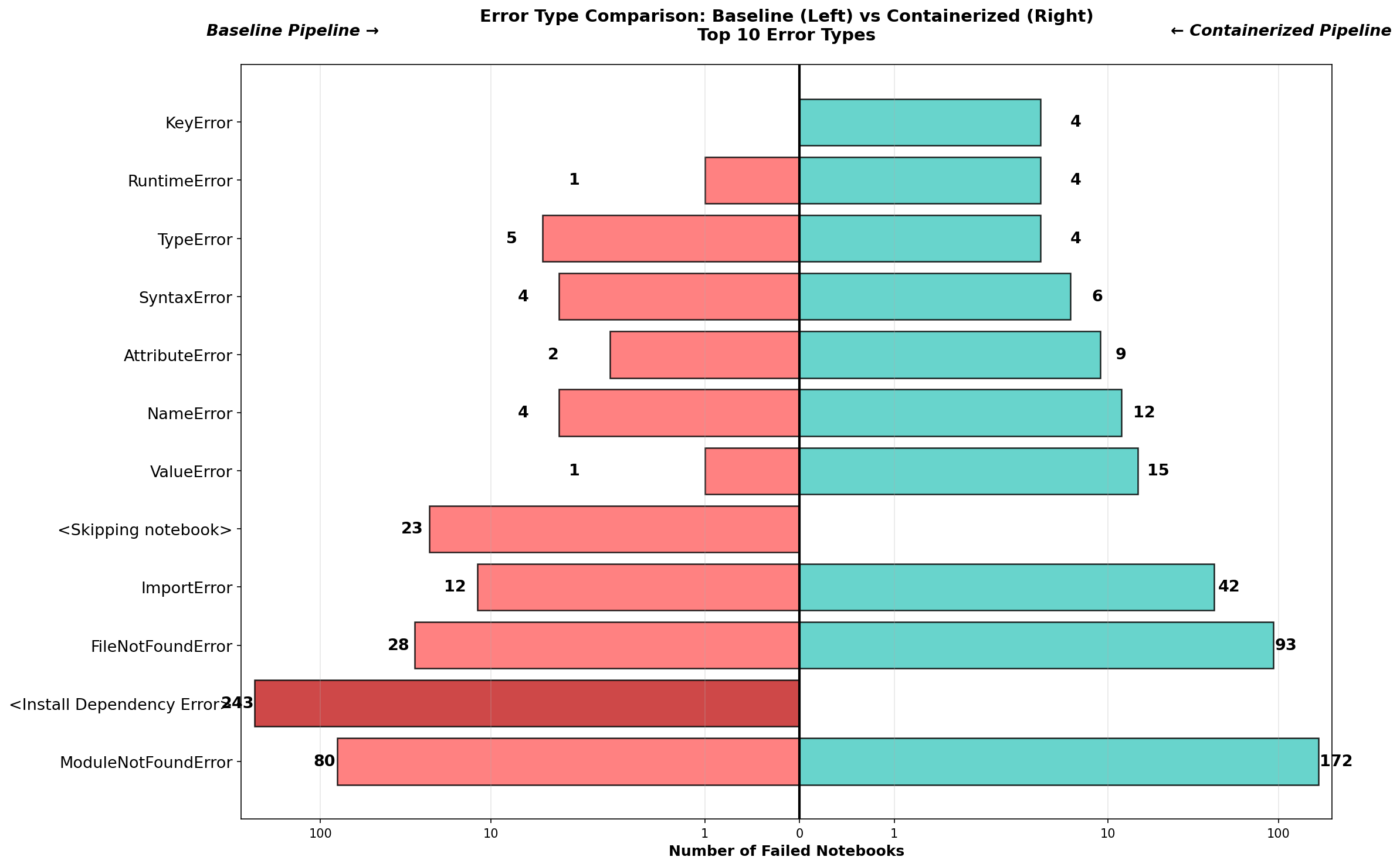}
\caption{Comparison of error types in baseline and containerized pipeline
}
\label{fig:error_comparison_base_container}
\vspace{-10pt}
\end{figure}

\textbf{Persistent Errors}
Figure \ref{fig:error_comparison_base_container} compares the distribution of error types observed in the baseline and containerized pipelines across 443  notebooks. In the baseline pipeline, failures are dominated by Install Dependency Error (243 cases, 59.7\%), followed by ModuleNotFoundError (80 cases, 19.7\%) and FileNotFoundError (26 cases, 6.4\%). This indicates that the primary bottleneck in the non-containerized setup is environment configuration and dependency resolution. In contrast, the containerized pipeline substantially reduces installation-related failures and shifts the error profile toward runtime exceptions. The most frequent errors under containerization are ModuleNotFoundError (172 cases, 44.8\%) and FileNotFoundError (93 cases, 24.2\%), followed by ImportError (42 cases, 10.9\%). This shift suggests that repository-level containerization successfully mitigates environment setup issues but exposes latent runtime assumptions, such as missing internal modules or unavailable data files of Class (ii) (Persistent Errors).

\begin{figure}
\centering
\includegraphics[width=\textwidth]{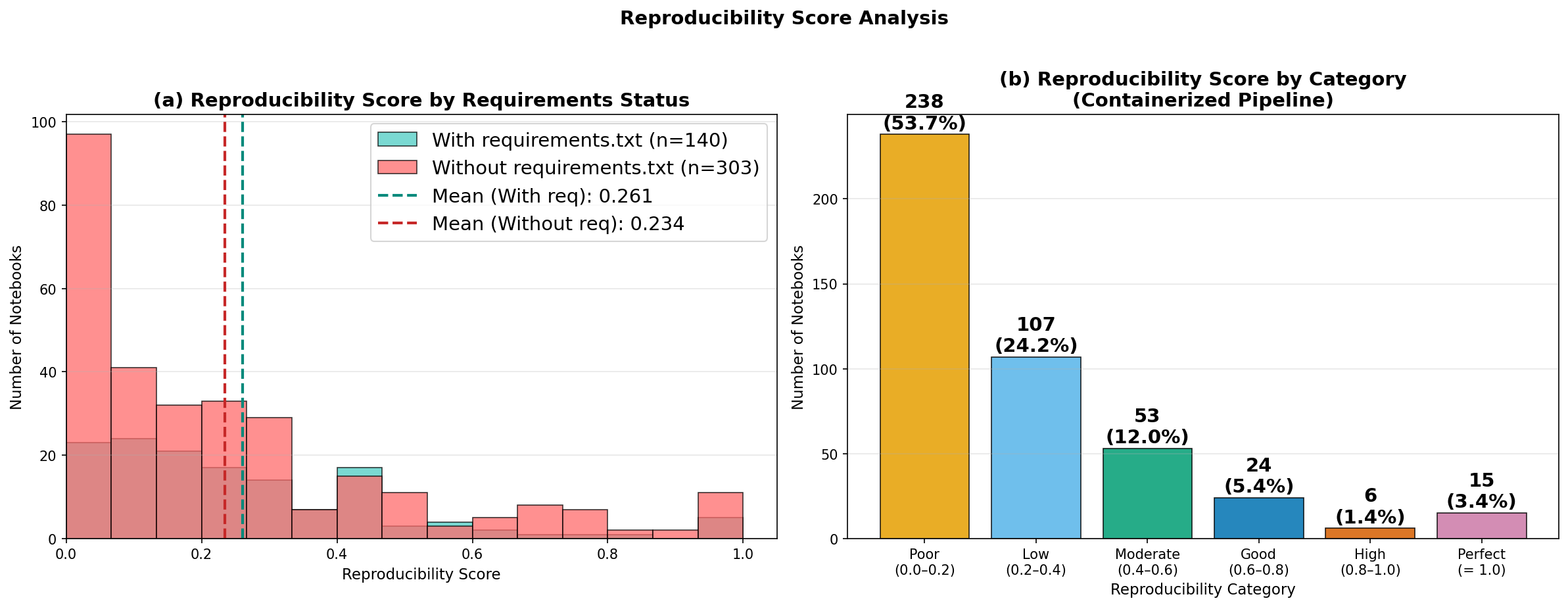}
\caption{Analysis of output fidelity for the containerized pipeline: (a) distribution of Reproducibility Scores for notebooks with vs. without \texttt{requirements.txt} files; (b) classification of notebooks by Reproducibility Score categories.
}
\label{fig:reproducibility_analysis_combined}
\vspace{-10pt}
\end{figure}

\textbf{Impact of Explicit Requirements} Figure \ref{fig:reproducibility_analysis_combined}(a) presents the distribution of reproducibility scores for notebooks 
that had
an explicit \texttt{requirements.txt} file in the original repository versus those that did not. Notebooks that include a dependency specification (n = 140) exhibit a slightly higher mean reproducibility score (0.261) compared to those without explicit requirements (n = 303; mean = 0.234). While both distributions are skewed toward lower reproducibility scores, indicating that full output equivalence remains challenging across environments, the presence of a \texttt{requirements.txt} file correlates with modestly improved consistency. 
Notably, notebooks without explicit dependency declarations show greater variance, including a higher concentration of very low reproducibility scores. This suggests that explicit environment specification contributes to more stable execution but is insufficient to guarantee deterministic results. Repository-level containerization mitigates some dependency-related issues; however, reproducibility remains influenced by runtime assumptions, external data dependencies, and non-deterministic computations. These findings indicate that while explicit environment specifications foster more stable execution, they are not a total remedy for Class (iii) Reproducibility Drift.

\begin{figure}
\centering
\includegraphics[width=0.75\textwidth, keepaspectratio]{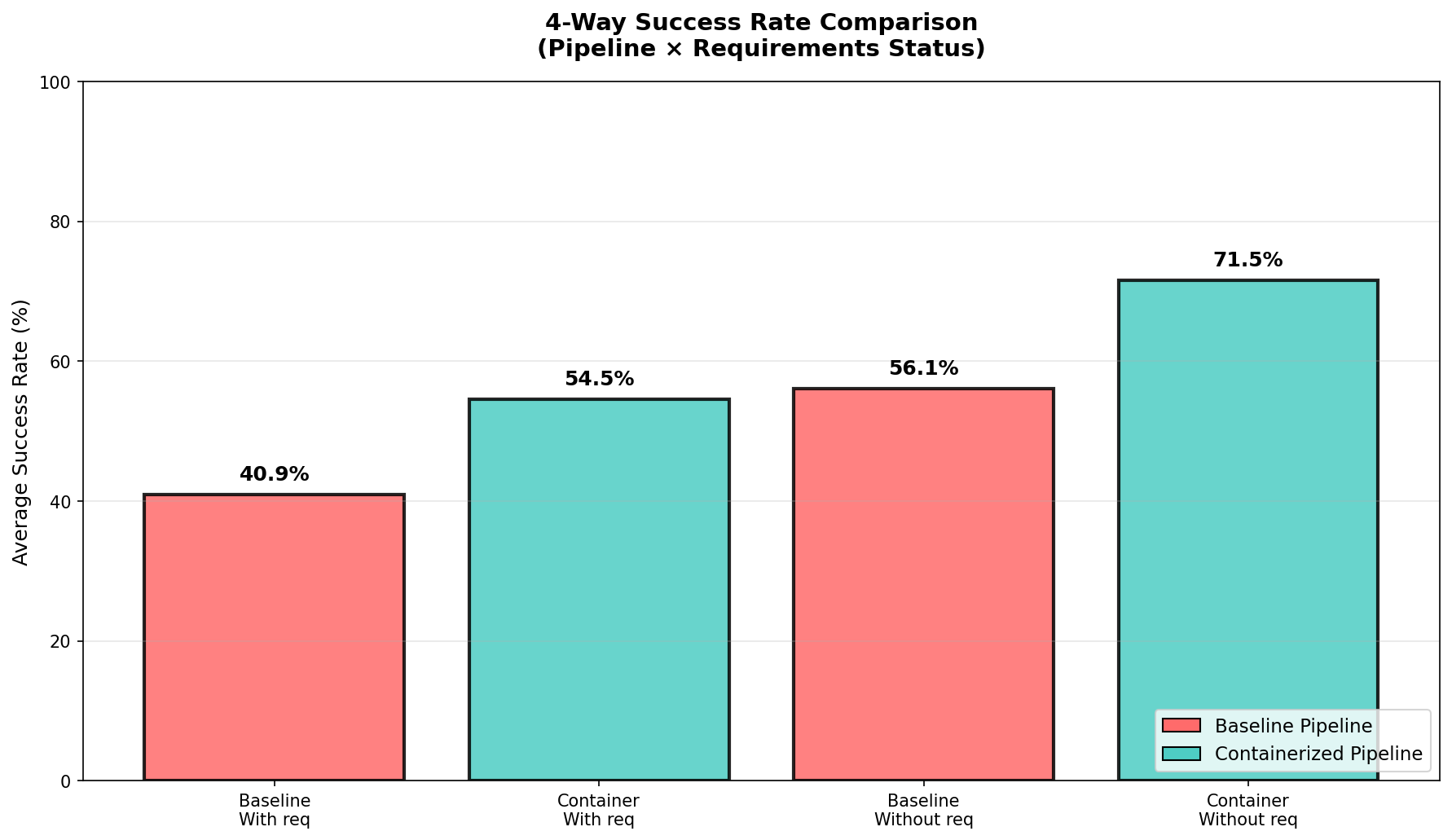}
\caption{Repository-level success rate (of reproducing the results of the original notebook) by requirements status, for both pipelines
}
\label{fig:four_way_comparison}
\vspace{-10pt}
\end{figure}

\textbf{Impact of Automated Environment Provisioning}
Figure \ref{fig:four_way_comparison} compares the Success Rate—defined as the percentage of notebooks that finished execution without runtime errors or kernel crashes—across the baseline and containerized pipelines. The repositories are categorized based on their original state in the PubMed Central database. To facilitate these results, our approach automatically generates and injects a \texttt{requirements.txt} file into every repository. The impact of this intervention is most visible in the ``Without req'' category, where the containerized pipeline achieved a 71.5\% success rate, demonstrating that standardized environment provisioning can effectively recover the execution integrity of undocumented research code.

\textbf{Reproducibility Gap} Figure \ref{fig:reproducibility_analysis_combined}(b) illustrates the distribution of the Reproducibility Score, defined as the ratio of identical output cells produced by the containerized pipeline to the total number of code cells in the original author’s notebook. Despite successful execution environments, a ``reproducibility gap'' exists, where 53.7\% (238 notebooks) fall into the Poor (0.0–0.2) category. This indicates that while the code executes, the specific results often diverge from the original outputs. Conversely, only 3.4\% (15 notebooks) achieved a Perfect score (1.0), representing a complete cell-for-cell match with the original repository.

A critical factor contributing to the ``reproducibility gap'' is the presence of non-deterministic operations within the dataset. Out of the 443 notebooks evaluated, 114 (25.7\%) were identified as containing non-deterministic patterns. We identify non-determinism using a set of static detection patterns targeting common sources of execution variability, including calls to \textit{random.*}, \textit{uuid.*}, \textit{np.random/numpy.random}, \textit{time.time}, \textit{datetime.now} and \textit{os.environ}. These patterns capture typical runtime behaviors that introduce variability across executions, such as stochastic sampling, time-dependent operations, environment-specific configuration and dynamically generated identifiers. The presence of such constructs explains a substantial portion of output differences observed even under containerized execution. 
The high prevalence of these patterns explains why a notebook can achieve a 100\% Success Rate (executing without errors) while yielding a poor Reproducibility Score. Even with perfectly provisioned environments and injected requirements, these functions ensure ``same cells'' rarely produce bit-wise identical output compared to the original author. This confirms that while containerization resolves environment-related failures (Class i), it unmasks deep-seated non-determinism and runtime dependencies characterizing Class (iii) Reproducibility Drift.

\input{Table1}
\par \textbf{Taxonomy of Execution Failures}
To satisfy \reqref{R4}, our structured logging and error classification mechanisms enabled us to group irreproducibility causes into four dominant failure categories. This taxonomy provides a granular engineering explanation for why repositories fall into Class (ii) or (iv):
\begin{itemize}
    \item \textbf{Dependency Obsolescence (Environment Decay)}: Failures often stem from outdated or unpinned libraries incompatible with modern environments. For instance, some repositories triggered a \textit{ModuleNotFoundError} during the container run phase, accounting for 44.8\% of containerized failures in Figure \ref{fig:error_comparison_base_container} due to deprecated scikit-learn APIs lacking version constraints.
    \item \textbf{Hardcoded or Missing Data Paths}: Many notebooks trigger a FileNotFoundError (24.2\% of containerized cases) by referencing absolute, machine-specific paths that do not exist within isolated environments.
    \item \textbf{External Credential Dependencies}: Some workflows (e.g., those requiring AWS or S3 access) depend on external credentials. These authentication requirements cannot be autonomously reconstructed in a clean-room setting.
    \item \textbf{Intrinsic Logical or Structural Errors}: Automated re-execution exposed pre-existing defects, such as \textit{SyntaxError}, \textit{NameError} or undefined variables, indicating that some artifacts were non-executable even in the original state.
\end{itemize}
Together, these categories demonstrate that irreproducibility arises from a combination of environment drift, implicit infrastructure assumptions, external service coupling, and latent code defects. 

\textbf{Data and code availability} 
We used
the dataset from \cite{samuel2023dataset}, also accessible as a knowledge graph \cite{samuel2024FAIR}.
The complete pipeline implementation is available\\ at \href{https://github.com/Sheeba-Samuel/computational-reproducibility-pmc-docker}{https://github.com/Sheeba-Samuel/computational-reproducibility-pmc-docker}.

\section{Discussion}
The results of our evaluation highlight both the potential and the inherent challenges of automating computational reproducibility assessments in the context of the scholarly record. Our findings underscore that while containerization is  powerful, it is not a ``silver bullet'' that can solve all forms of irreproducibility.

\par{\textbf{Minimizing Environmental Decay:}}
The primary strength of our pipeline lies in its ability to contain environmental decay rather than eliminate it entirely. By abstracting execution environments into Docker containers, we mitigate the “it works on my machine” problem and shield notebooks from host-level inconsistencies. The results show that many artifacts previously deemed “broken” due to dependency conflicts or operating system mismatches can be restored through automated environment reconstruction. This indicates that a substantial portion of observed irreproducibility stems not from flawed research logic, but from inadequate environment management—suggesting that parts of the so-called reproducibility crisis are, in practice, an infrastructure and dependency management problem.

\par{\textbf{Implications for Scholarly Communication}}
The proposed tool can support multiple stakeholders across the scholarly ecosystem by embedding automated reproducibility assessment into different stages of the research lifecycle.
For researchers, it can serve as a pre-publication validation mechanism to identify missing dependencies, non-deterministic code, implicit assumptions, and undocumented data requirements before submission.
Peer reviewers can leverage structured execution reports and reproducibility scores for a systematic and time-efficient assessment of computational robustness.
Editorial workflows can replace manual checklists with automated evidence-based reproducibility checks.
Organizations conducting large-scale reproducibility studies can use the pipeline to track reproducibility trends and infrastructure decay across various venues, domains and publication years.

\par{\textbf{Limitations and Challenges}}
Despite the robustness of the Docker-based approach, several limitations remain. In repositories without explicit dependency specifications, import analysis captures only package names but not versions, which may lead to compatibility issues. Although a small number of repositories provide their own Dockerfiles, these were not incorporated into the current pipeline (this remains future work). Because the system prioritizes automation and scalability, some failures stem from missing external resources or undocumented assumptions that cannot be resolved automatically. Containerization also cannot prevent link rot, expired API keys, or disappearing cloud resources, making externally dependent notebooks inherently fragile unless paired with persistent archives such as Zenodo. Furthermore, the standardized CPU-based infrastructure limits support for GPU-dependent notebooks, and reliance on a modern Python base image creates challenges for legacy Python 2.7 artifacts. Finally, stochastic non-determinism—observed in 25.7\% of notebooks—means that successful execution does not guarantee identical outputs. Overall, the pipeline advances proactive reproducibility engineering by systematically mitigating environment drift while transparently exposing structural limitations of computational research artifacts.
\vspace{-10pt}
\section{Conclusion}
This paper presented a repository-level containerization pipeline for the automated analysis of scholarly Jupyter notebooks at scale. Framed from a Web Engineering perspective, our contribution is the design of an end-to-end process integrating repository acquisition, automated environment reconstruction, containerized execution, and fine-grained reproducibility analysis. Across 116 repositories and 443 notebooks, results show containerization substantially reduces dependency installation failures—the dominant issue in our baseline—shifting failure modes toward application-level errors. Since 114 notebooks exhibited non-determinism, environment isolation alone proved insufficient without controlling stochastic behavior. Ultimately, repository-level encapsulation improves fault localization, transparency, and diagnostics beyond binary metrics.

The key contribution is a scalable process architecture treating research artifacts as deployable web resources subject to drift and incomplete specifications. By mitigating dependency decay, the pipeline preserves long-term executability. Future work will evolve the system into a more intelligent framework, integrating LLMs to analyze code comments and logs for automated dependency prediction and error fixing. We plan to scale the pipeline to the entire dataset \cite{samuel2023dataset} and map results into the FAIR Jupyter Knowledge Graph \cite{samuel2024FAIR}.
We will extend support to HPC environments via NVIDIA Docker and Singularity for GPU-intensive workloads and apply this logic to the entire dataset for longitudinal analysis. We aim to integrate this pipeline into manuscript submission workflows for an automated `Reproducibility Badge'' system. Finally, future iterations will explore normalization techniques to manage stochastic non-determinism, bridging the gap between execution success and numerical identity.
\begin{credits}
\subsubsection{\ackname} 
This work was supported in part by the German Research Foundation (DFG) through the following projects:
Jupyter4NFDI (\href{https://gepris.dfg.de/gepris/projekt/521453681}{DFG 521453681}  \cite{hagemeier2025Jupyter4NFDI}), 
find.software (\href{https://gepris.dfg.de/gepris/projekt/567156310}{DFG 567156310} \cite{gey2025find.software}), MaRDI (\href{https://gepris.dfg.de/gepris/projekt/460135501}{DFG 460135501} \cite{the_mardi_consortium_2022_6552436} as well as
SeDOA (\href{https://gepris.dfg.de/gepris/projekt/556323977}{DFG 556323977}  \cite{stacker2025SeDOA}) and HYP*MOL (\href{https://gepris.dfg.de/gepris/projekt/514664767}{DFG 514664767}). 
The text of this manuscript was improved with the following AI tools: ChatGPT and Gemini.
\end{credits}
%
%
%
\bibliographystyle{splncs04}
\bibliography{main}
\end{document}

%% file: Table1.tex
\markboth{}{}
\thispagestyle{headings}
{

\footnotesize 
\begin{sidewaystable}[t]

\centering
\caption{Comparison between the previous and the present study across dependency installation, execution outcomes and reproducibility metrics.
}
\label{tab:prev-vs-this}
\setlength{\tabcolsep}{5pt}
\renewcommand{\arraystretch}{1.2}

\begin{tabular}{p{3.5cm} c cc cc cc cc }
\toprule
\multirow{2}{*}{\textbf{Repository (ID)}} &
\multirow{2}{*}{\textbf{NB ID}} &
\multicolumn{2}{c}{\textbf{Dependency Install}} &
\multicolumn{2}{c}{\textbf{Execution Status}} &
\multicolumn{2}{c}{\textbf{Diff Cells}} &
\multicolumn{2}{c}{\textbf{Duration (s)}} \\
\cmidrule(lr){3-4}
\cmidrule(lr){5-6}
\cmidrule(lr){7-8}
\cmidrule(lr){9-10}

& &
\textbf{Prev.} & \textbf{This} &
\textbf{Prev.} & \textbf{This} &
\textbf{Prev.} & \textbf{This} &
\textbf{Prev.} & \textbf{This} \\
\midrule

\hline
\textbf{A. Environment Failures Resolved} \\
\href{https://reproduceme.uni-jena.de/#/dataset/fairjupyter/query?query=SELECT%20%3Frepository%20%3Fp%20%3Fo%0AWHERE%20%7B%20%20%0A%20%20%3Chttps%3A%2F%2Fw3id.org%2Freproduceme%2Frepository_16%3E%20%3Fp%20%3Fo%20.%0A%7D%20%20}{gpax (16)} &
\href{https://reproduceme.uni-jena.de/#/dataset/fairjupyter/query?query=SELECT%20%3Fp%20%3Fo%0AWHERE%20%7B%20%20%0A%20%20%3Chttps%3A%2F%2Fw3id.org%2Freproduceme%2Fnotebook_32%3E%20%3Fp%20%3Fo%20.%0A%7D%20%20}{32} &
Fail & Success &
Fail & Success &
- & 4 &
- & 89.0 \\

\href{https://reproduceme.uni-jena.de/#/dataset/fairjupyter/query?query=SELECT%20%3Frepository%20%3Fp%20%3Fo%0AWHERE%20%7B%20%20%0A%20%20%3Chttps%3A%2F%2Fw3id.org%2Freproduceme%2Frepository_27%3E%20%3Fp%20%3Fo%20.%0A%7D%20%20}{mofax (27)} & 
\href{https://reproduceme.uni-jena.de/#/dataset/fairjupyter/query?query=SELECT%20%3Fp%20%3Fo%0AWHERE%20%7B%20%20%0A%20%20%3Chttps%3A%2F%2Fw3id.org%2Freproduceme%2Fnotebook_32%3E%20%3Fp%20%3Fo%20.%0A%7D%20%20}{83} & 
Fail & Success &
Install Dependency Error & Success &
- & 30 &
- & 108.0 \\

\href{https://reproduceme.uni-jena.de/#/dataset/fairjupyter/query?query=SELECT%20%3Frepository%20%3Fp%20%3Fo%0AWHERE%20%7B%20%20%0A%20%20%3Chttps%3A%2F%2Fw3id.org%2Freproduceme%2Frepository_22%3E%20%3Fp%20%3Fo%20.%0A%7D%20%20}{PhysiCOOL (22)}  &
\href{https://reproduceme.uni-jena.de/#/dataset/fairjupyter/query?query=SELECT%20%3Fp%20%3Fo%0AWHERE%20%7B%20%20%0A%20%20%3Chttps%3A%2F%2Fw3id.org%2Freproduceme%2Fnotebook_53%3E%20%3Fp%20%3Fo%20.%0A%7D%20%20}{53} &
Fail & Success &
Install Dependency Error & ModuleNotFoundError &
- & 4 &
- & 4\\

\href{https://reproduceme.uni-jena.de/#/dataset/fairjupyter/query?query=SELECT%20%3Frepository%20%3Fp%20%3Fo%0AWHERE%20%7B%20%20%0A%20%20%3Chttps%3A%2F%2Fw3id.org%2Freproduceme%2Frepository_9%3E%20%3Fp%20%3Fo%20.%0A%7D%20%20}{pymCADRE (10)} &
\href{https://reproduceme.uni-jena.de/#/dataset/fairjupyter/query?query=SELECT%20%3Fp%20%3Fo%0AWHERE%20%7B%20%20%0A%20%20%3Chttps%3A%2F%2Fw3id.org%2Freproduceme%2Fnotebook_13%3E%20%3Fp%20%3Fo%20.%0A%7D%20%20}{13} &
Fail & Success &
Install Dependency Error & PermissionError &
- & 9 &
- & 6.0 \\

\hline

\textbf{B. Persistent Errors} \\
\href{https://reproduceme.uni-jena.de/#/dataset/fairjupyter/query?query=SELECT%20%3Frepository%20%3Fp%20%3Fo%0AWHERE%20%7B%20%20%0A%20%20%3Chttps%3A%2F%2Fw3id.org%2Freproduceme%2Frepository_3%3E%20%3Fp%20%3Fo%20.%0A%7D%20%20}{elastic-blast-demos (3)} &
\href{https://reproduceme.uni-jena.de/#/dataset/fairjupyter/query?query=SELECT%20%3Fp%20%3Fo%0AWHERE%20%7B%20%20%0A%20%20%3Chttps%3A%2F%2Fw3id.org%2Freproduceme%2Fnotebook_1%3E%20%3Fp%20%3Fo%20.%0A%7D%20%20}{1} &
Success & Success &
SyntaxError & SyntaxError &
0 & 20 &
3.73 & 35 \\

\href{https://reproduceme.uni-jena.de/#/dataset/fairjupyter/query?query=SELECT%20%3Frepository%20%3Fp%20%3Fo%0AWHERE%20%7B%20%20%0A%20%20%3Chttps%3A%2F%2Fw3id.org%2Freproduceme%2Frepository_8%3E%20%3Fp%20%3Fo%20.%0A%7D%20%20}{braincharts (8)} & 
\href{https://reproduceme.uni-jena.de/#/dataset/fairjupyter/query?query=SELECT%20%3Fp%20%3Fo%0AWHERE%20%7B%20%20%0A%20%20%3Chttps%3A%2F%2Fw3id.org%2Freproduceme%2Fnotebook_3%3E%20%3Fp%20%3Fo%20.%0A%7D%20%20}{3} &
Success & Success &
FileNotFoundError & FileNotFoundError &
0 & 20 &
80.4 & 358.0 \\

\href{https://reproduceme.uni-jena.de/#/dataset/fairjupyter/query?query=SELECT%20%3Frepository%20%3Fp%20%3Fo%0AWHERE%20%7B%20%20%0A%20%20%3Chttps%3A%2F%2Fw3id.org%2Freproduceme%2Frepository_17%3E%20%3Fp%20%3Fo%20.%0A%7D%20%20}{eubucco (17)} & 
\href{https://reproduceme.uni-jena.de/#/dataset/fairjupyter/query?query=SELECT%20%3Fp%20%3Fo%0AWHERE%20%7B%20%20%0A%20%20%3Chttps%3A%2F%2Fw3id.org%2Freproduceme%2Fnotebook_43%3E%20%3Fp%20%3Fo%20.%0A%7D%20%20}{43} &
Sucess & Success &
File Not Found Error & File Not Found Error &
0 & 2 &
2.8 & 3.0 \\

\hline

\textbf{C. Reproducibility Drift} \\
\href{https://reproduceme.uni-jena.de/#/dataset/fairjupyter/query?query=SELECT%20%3Frepository%20%3Fp%20%3Fo%0AWHERE%20%7B%20%20%0A%20%20%3Chttps%3A%2F%2Fw3id.org%2Freproduceme%2Frepository_15%3E%20%3Fp%20%3Fo%20.%0A%7D%20%20}{Reuse\_in\_processes (15)} &
\href{https://reproduceme.uni-jena.de/#/dataset/fairjupyter/query?query=SELECT%20%3Fp%20%3Fo%0AWHERE%20%7B%20%20%0A%20%20%3Chttps%3A%2F%2Fw3id.org%2Freproduceme%2Fnotebook_30%3E%20%3Fp%20%3Fo%20.%0A%7D%20%20}{30} &
Success & Success &
Success & Success &
0 & 3 &
4.6 & 8.0 \\

\href{https://reproduceme.uni-jena.de/#/dataset/fairjupyter/query?query=SELECT%20%3Frepository%20%3Fp%20%3Fo%0AWHERE%20%7B%20%20%0A%20%20%3Chttps%3A%2F%2Fw3id.org%2Freproduceme%2Frepository_5005%3E%20%3Fp%20%3Fo%20.%0A%7D%20%20}{idr-notebooks (5005)} &
\href{https://reproduceme.uni-jena.de/#/dataset/fairjupyter/query?query=SELECT%20%3Fp%20%3Fo%0AWHERE%20%7B%20%20%0A%20%20%3Chttps%3A%2F%2Fw3id.org%2Freproduceme%2Fnotebook_26310%3E%20%3Fp%20%3Fo%20.%0A%7D%20%20}{26310} &
Success & Success &
Success & Success &
1 & 1 &
1.0 & 4.0 \\
\hline

\textbf{D. Regressions} \\
\href{https://reproduceme.uni-jena.de/#/dataset/fairjupyter/query?query=SELECT%20%3Frepository%20%3Fp%20%3Fo%0AWHERE%20%7B%20%20%0A%20%20%3Chttps%3A%2F%2Fw3id.org%2Freproduceme%2Frepository_9%3E%20%3Fp%20%3Fo%20.%0A%7D%20%20}{evidence\_embracing\_nm (9)} &
\href{https://reproduceme.uni-jena.de/#/dataset/fairjupyter/query?query=SELECT%20%3Fp%20%3Fo%0AWHERE%20%7B%20%20%0A%20%20%3Chttps%3A%2F%2Fw3id.org%2Freproduceme%2Fnotebook_9%3E%20%3Fp%20%3Fo%20.%0A%7D%20%20}{9} &
Success & Invalid url &
ModuleNotFoundError & - &
0 & - &
2.21 & - \\

\href{https://reproduceme.uni-jena.de/#/dataset/fairjupyter/query?query=SELECT%20%3Frepository%20%3Fp%20%3Fo%0AWHERE%20%7B%20%20%0A%20%20%3Chttps%3A%2F%2Fw3id.org%2Freproduceme%2Frepository_196%3E%20%3Fp%20%3Fo%20.%0A%7D%20%20}{omero-guide-python (196)} &
\href{https://reproduceme.uni-jena.de/#/dataset/fairjupyter/query?query=SELECT%20%3Fp%20%3Fo%0AWHERE%20%7B%20%20%0A%20%20%3Chttps%3A%2F%2Fw3id.org%2Freproduceme%2Fnotebook_718%3E%20%3Fp%20%3Fo%20.%0A%7D%20%20}{718} &
Fail & Docker build Fail &
Install Dependency Error & Kernel not found &
- & - &
-  & - \\

\href{https://reproduceme.uni-jena.de/#/dataset/fairjupyter/query?query=SELECT%20%3Frepository%20%3Fp%20%3Fo%0AWHERE%20%7B%20%20%0A%20%20%3Chttps%3A%2F%2Fw3id.org%2Freproduceme%2Frepository_36%3E%20%3Fp%20%3Fo%20.%0A%7D%20%20}{2022.CC (36)} &
\href{https://reproduceme.uni-jena.de/#/dataset/fairjupyter/query?query=SELECT%20%3Fp%20%3Fo%0AWHERE%20%7B%20%20%0A%20%20%3Chttps%3A%2F%2Fw3id.org%2Freproduceme%2Fnotebook_99%3E%20%3Fp%20%3Fo%20.%0A%7D%20%20}{99} &
Success & Success &
Sucess & Success &
0 & 5 &
6.6 & 6.0 \\

\href{https://reproduceme.uni-jena.de/#/dataset/fairjupyter/query?query=SELECT%20%3Frepository%20%3Fp%20%3Fo%0AWHERE%20%7B%20%20%0A%20%20%3Chttps%3A%2F%2Fw3id.org%2Freproduceme%2Frepository_17%3E%20%3Fp%20%3Fo%20.%0A%7D%20%20}{eubucco (17)} &
\href{https://reproduceme.uni-jena.de/#/dataset/fairjupyter/query?query=SELECT%20%3Fp%20%3Fo%0AWHERE%20%7B%20%20%0A%20%20%3Chttps%3A%2F%2Fw3id.org%2Freproduceme%2Fnotebook_42%3E%20%3Fp%20%3Fo%20.%0A%7D%20%20}{42} &
Sucess & Success &
<Skipping notebook> & TypeError &
- & 10 &
- & 12 \\

\href{https://reproduceme.uni-jena.de/#/dataset/fairjupyter/query?query=SELECT%20%3Frepository%20%3Fp%20%3Fo%0AWHERE%20%7B%20%20%0A%20%20%3Chttps%3A%2F%2Fw3id.org%2Freproduceme%2Frepository_174%3E%20%3Fp%20%3Fo%20.%0A%7D%20%20}{Minimal\_Cell (174)}  &
\href{https://reproduceme.uni-jena.de/#/dataset/fairjupyter/query?query=SELECT%20%3Fp%20%3Fo%0AWHERE%20%7B%20%20%0A%20%20%3Chttps%3A%2F%2Fw3id.org%2Freproduceme%2Fnotebook_53%3E%20%3Fp%20%3Fo%20.%0A%7D%20%20}{53} &
Fail & Docker Build Fail &
Install Dependency Error & Module Not Found &
- & - &
- & -\\

\href{https://reproduceme.uni-jena.de/#/dataset/fairjupyter/query?query=SELECT%20%3Frepository%20%3Fp%20%3Fo%0AWHERE%20%7B%20%20%0A%20%20%3Chttps%3A%2F%2Fw3id.org%2Freproduceme%2Frepository_138%3E%20%3Fp%20%3Fo%20.%0A%7D%20%20}{imos-user-code-library (138)} &
\href{https://reproduceme.uni-jena.de/#/dataset/fairjupyter/query?query=SELECT%20%3Fp%20%3Fo%0AWHERE%20%7B%20%20%0A%20%20%3Chttps%3A%2F%2Fw3id.org%2Freproduceme%2Fnotebook_530%3E%20%3Fp%20%3Fo%20.%0A%7D%20%20}{530} &
Success & Success &
ModuleNotFoundError & Notebook not found &
1 & - &
1.04 & -  \\


\bottomrule
\end{tabular}

\vspace{0.3em}
\footnotesize{
\textbf{Prev.} refers to results reported in the original study.
\textbf{This} refers to results obtained using our container-based reproducibility pipeline. \textbf{NB ID} denotes the unique identifier for each notebook, while \textbf{Duration} represents its total execution time. 
\textbf{Dependency Install} denotes the success or failure of the environment reconstruction phase. \textbf{Execution Status} denotes the outcome of the notebook’s runtime phase once the environment is established.
The provided links direct to the SPARQL endpoint of the FAIR Jupyter Knowledge Graph \cite{samuel2024FAIR}.
}
\end{sidewaystable}
}